\documentclass[sigconf]{acmart}

\usepackage{booktabs}
\usepackage{multirow}
\usepackage{float}

\usepackage{graphicx}
\usepackage{float}
\usepackage{placeins}
\usepackage{float} 
\usepackage{algorithm}
\usepackage{algpseudocode}

\usepackage{listings}

\hypersetup{hidelinks}

\renewcommand\footnotetextcopyrightpermission[1]{}

\AtBeginDocument{%
  \providecommand\BibTeX{{%
    \normalfont B\kern-0.5em{\scshape i\kern-0.25em b}\kern-0.8em\TeX}}}

\graphicspath{{./images/}} 
\begin{document}

\title{ Causally Disentangled Contrastive Learning for Multilingual Speaker Embeddings}

\author{
  Mariëtte Olijslager$^{1}$, 
  Seyed Sahand Mohammadi Ziabari$^{1,2}$, 
  Ali Mohammed Mansoor Alsahag$^{1}$
}

\affiliation{%
  \institution{$^{1}$Informatics Institute, University of Amsterdam}
  \streetaddress{1098XH Science Park}
  \city{Amsterdam}
  \country{The Netherlands}
}

\affiliation{%
  \institution{$^{2}$Department of Computer Science and Technology, SUNY Empire State University}
  \city{Saratoga Springs, NY}
  \country{USA}
}

\email{mariette.olijslager@student.uva.nl, sahand.ziabari@sunyempire.edu, a.m.m.alsahag@uva.nl }

\begin{abstract}
Self-supervised speaker embeddings are widely used in speaker verification systems, but prior work has shown that they often encode sensitive demographic attributes, raising fairness and privacy concerns. This paper investigates the extent to which demographic information—specifically, gender, age, and accent—is present in SimCLR-trained speaker embeddings and whether such leakage can be mitigated without severely degrading speaker verification performance. We study two debiasing strategies: adversarial training through gradient reversal and a causal bottleneck architecture that explicitly separates demographic and residual information. Demographic leakage is quantified using both linear and nonlinear probing classifiers, while speaker verification performance is evaluated using ROC-AUC and EER. Our results show that gender information is strongly and linearly encoded in baseline embeddings, whereas age and accent are weaker and primarily nonlinearly represented. Adversarial debiasing reduces gender leakage, but has limited effect on age and accent, and introduces a clear trade-off with verification accuracy. The causal bottleneck further suppresses demographic information, particularly in the residual representation, but incurs substantial performance degradation. These findings highlight fundamental limitations in mitigating demographic leakage in self-supervised speaker embeddings and clarify the trade-offs inherent in current debiasing approaches.
\end{abstract}

\keywords{Speaker verification,
Self-supervised learning,
Contrastive learning,
Demographic leakage,
Adversarial debiasing}

\fancyhead{}
\maketitle

\section{Introduction}
\label{sec:introduction}
In recent years, deep learning techniques have significantly advanced the field of speaker recognition and verification. Large-scale models trained on extensive speech corpora are now capable of capturing highly discriminative acoustic representations, enabling strong speaker identification even under challenging acoustic conditions \citep{snyder2018xvectors,desplanques2020ecapa}. However, as these models become increasingly powerful, concerns have emerged regarding the unintentional encoding of sensitive demographic information, such as gender, age, or accent, within their latent representations. This phenomenon, known as \textit{demographic leakage}, poses potential risks to fairness, privacy, and ethical compliance. It may lead to systematic biases across demographic groups or enable unauthorized inference of protected attributes from supposedly anonymized embeddings \citep{raj2019probing,luu2022disentangled}. Addressing such leakage is therefore essential for ensuring fair and responsible deployment of speech technologies \citep{kusner2017counterfactual}. Importantly, such leakage can arise not only from direct statistical correlations, but also from underlying causal mechanisms that link demographic attributes to learned representations, motivating approaches that go beyond correlation-based mitigation.

Speaker embeddings have historically been learned using supervised classification objectives, which require large amounts of labeled speaker data and can inadvertently encode spurious correlations with non-speaker attributes. To mitigate these limitations, recent work has proposed self-supervised contrastive learning approaches such as SimCLR and wav2vec. These methods demonstrate that meaningful speech representations can be learned by maximizing agreement between augmented views of the same input, without requiring speaker labels \citep{baevski2020wav2vec}. Despite their scalability and representational strength, contrastive learning methods do not inherently enforce invariance with respect to sensitive demographic attributes. As a result, demographic information may persist in the learned embeddings even when it is not required for downstream speaker verification, potentially compromising fairness and privacy \citep{li2023embeddingleak}. This motivates a structured investigation into how demographic leakage manifests in contrastive speaker embeddings and how it can be mitigated.

Despite the growing evidence of demographic leakage in speaker representations, relatively few studies have systematically quantified or mitigated this bias in self-supervised contrastive embeddings. Most prior work has focused on adversarial or multi-task debiasing strategies in supervised speaker recognition, revealing inherent fairness–utility trade-offs and limited ability to fully suppress demographic information \citep{peri2022bias,peri2023study}. Architectural adaptations aimed at reducing demographic disparities have also been proposed in supervised settings \citep{shen2022group}. However, these approaches are often not directly applicable to self-supervised contrastive representations, which differ fundamentally in their learning objectives and inductive biases. In contrastive learning, the objective encourages representations that are maximally discriminative across utterances, implicitly reinforcing any demographic regularities present in the data and leading to entangled embeddings in which task-relevant and sensitive attributes are jointly encoded. As a result, demographic leakage becomes a representation-learning problem rather than a downstream policy issue. In particular, the use of explicitly structured or causally motivated bottleneck architectures to isolate and control demographic information remains largely unexplored in self-supervised speaker embedding models.

This paper explores a comparative framework for mitigating demographic leakage in self-supervised speaker embeddings. First, contrastive learning method SimCLR is used to create a baseline that is trained to learn speaker-discriminative representations. Second, adversarial debiasing is applied to suppress demographic information, and leakage is evaluated using both linear and nonlinear probing classifiers to assess demographic leakage information beyond linear separability. Third, a causal bottleneck layer is introduced at the end of the representation pipeline, explicitly constraining the information flow from sensitive attributes into the final embedding. By modeling demographic attributes as causally upstream factors and restricting their influence on the learned representation, the causal bottleneck provides a novel alternative to just using adversarial objectives for mitigating demographic leakage.

This paper investigates demographic information leakage in self-supervised speaker embeddings and evaluates whether such leakage can be mitigated through adversarial debiasing and causal bottleneck architectures, while preserving speaker verification performance. The study first analyzes how SimCLR-trained speaker embeddings encode demographic attributes related to gender, age, and accent, using both linear and nonlinear probing classifiers to quantify the extent and structure of the leakage. It then examines the effectiveness of adversarial debiasing in suppressing demographic information and assesses how this intervention affects both linearly and nonlinearly encoded attributes as well as speaker verification accuracy. Finally, the paper evaluates whether introducing a causal bottleneck layer provides additional suppression of demographic leakage beyond adversarial debiasing alone, particularly for nonlinear demographic information, and analyzes the resulting trade-offs with speaker verification performance.

Through comprehensive evaluation, including speaker verification accuracy, demographic probing, and fairness-related metrics, this paper aims to characterize the effectiveness and limitations of different debiasing strategies for self-supervised speaker representations. By contrasting adversarial objectives with causally motivated architectural constraints, the work contributes to a deeper understanding of how demographic leakage arises and how it can be mitigated in modern contrastive learning frameworks.

\section{Related Work}
\label{sec:related_work}
This section reviews prior work related to demographic information leakage in speech representations and existing mitigation strategies. We first summarize evidence that modern speaker embedding models encode sensitive demographic attributes such as gender, age, and accent. We then review contrastive learning approaches in speech and fairness-aware representation learning, followed by causal and adversarial methods for disentangling protected attributes from learned embeddings. This progression motivates the methodological choices of the present study and highlights gaps that our work aims to address.

\subsection{Demographic information in speaker embeddings}

Understanding how speaker embeddings encode demographic information is crucial for designing fair and privacy-aware speaker verification systems. Recent work has investigated the extent to which popular embedding models retain gender, age, and accent cues, highlighting potential demographic leakage. Both WavLM~\cite{wavlm-demographic} (a self-supervised speech model) and ECAPA-TDNN~\cite{desplanques2020ecapa} (a supervised speaker-verification encoder) produce dense embeddings that preserve demographic cues, but neither architecture is designed to suppress such information and thus they are not considered state-of-the-art for mitigating demographic leakage. Prior analyses have quantified how different demographic attributes are encoded in these embeddings. Gender is almost linearly separable: a simple linear probe reaches over 99.8\% accuracy, indicating that a single dominant direction in the embedding space carries the gender signal. Age is encoded more diffusely, with regression probes achieving a mean absolute error of roughly 4.9 years, showing that age information is distributed across many dimensions and requires nonlinear mappings (e.g., multilayer perceptrons or ridge regression) for effective recovery \cite{wavlm-demographic}. Accent lies between these extremes: classification benchmarks such as CommonAccent demonstrate strong performance, yet consistently lower than gender prediction, indicating that accent cues are present but spread across multiple dimensions.  

These findings establish a consistent pattern: gender is the easiest, age the hardest, and accent intermediate to predict from speaker embeddings. Because these encoders are deep and nonlinear, the latent representations themselves are more complex; while gender can be read out with a linear classifier, age and accent often require nonlinear probing. Importantly, this nonlinear leakage motivates evaluating both linear and nonlinear probes in downstream experiments to fully quantify demographic information in embeddings.

Overall, the analyses highlight that raw WavLM or ECAPA-TDNN embeddings are entangled with respect to demographics. Achieving genuine privacy or fairness therefore requires additional objectives, such as adversarial debiasing, mutual-information minimization (as in WavShape~\cite{baser2025wavshape}), or contrastive designs that enforce invariance. This motivates the current study of contrastive self-supervised speaker embeddings and the introduction of causal bottleneck architectures, which explicitly aim to isolate speaker-discriminative information from demographic factors.

\subsection{Contrastive learning methods for speaker embeddings}

FairASR~\cite{fairasr2023} demonstrates that adversarial demographic suppression can reduce demographic gaps in ASR outputs while maintaining overall word-error rate (WER) comparable to baseline models. However, it relies on demographic labels during pre-training, reports only aggregate disparity reductions without detailed per-group statistics, and is evaluated mainly on in-domain corpora. Subsequent work highlights that fairness in ASR is context-dependent and that improvements on one demographic axis may not generalize across datasets or under distribution shift~\cite{fairness_nosizefitsall}. A recent contrastive method pairs gender-swapped utterances to produce gender-invariant embeddings, achieving strong cross-domain fairness while requiring minimal WER compromise~\cite{gender_swapped_contrastive}.

Despite these advances, prior approaches are limited in scope: they typically target a single demographic attribute, focus on ASR rather than speaker verification, and do not systematically evaluate nonlinear demographic leakage or intersectional effects.

In contrast, this paper investigates self-supervised speaker embeddings (e.g., SimCLR-based) and combines adversarial debiasing with a causal bottleneck to mitigate leakage across multiple demographic axes (gender, age, accent) and their intersections. Leakage is quantified using both linear and nonlinear probes, with bootstrapped confidence intervals to provide rigorous uncertainty estimates. Unlike FairASR and related ASR-focused methods, our work applies adversarial debiasing and causal bottlenecks specifically to speaker embeddings and systematically investigates nonlinear leakage and intersectional effects.

\subsection{Causal disentanglement}

To further suppress unwanted demographic information in learned representations, adversarial debiasing techniques can be incorporated through a causal bottleneck layer. The underlying idea, introduced in the Domain-Adversarial Neural Network (DANN)~\cite{ganin2016domain}, couples a main encoder with an auxiliary adversary trained to predict a domain or protected attribute from latent features, while a gradient-reversal layer forces the encoder to remove such information.

Subsequent work by Elazar and Goldberg~\cite{elazar2018adversarial} extended this concept to textual data, demonstrating that adversarial objectives reduce, but do not entirely eliminate, demographic leakage, highlighting the importance of verifying residual bias using independent attacker classifiers. Recent work has emphasized causal and representation-level approaches to bias mitigation. Koçadağ et al.~\cite{kocadag2026intersectional} propose an SCM-based framework for intersectional bias mitigation in word embeddings, showing that explicitly modeling causal relationships improves coherence and fairness. Similarly, Zhu et al.~\cite{zhu2025taskadaptive} introduce task-adaptive debiasing using structural causal models, demonstrating inherent trade-offs between task performance and bias mitigation.

Beyond causal modeling, Garbat et al.~\cite{garbat2026warmth} study debiasing of contextual embeddings along socially grounded dimensions, while Naranbat et al.~\cite{naranbat2025fairness} show that fairness outcomes are highly sensitive to evaluation design and domain choice.

\subsection{Debiasing representations via adversarial debiasing and bottlenecks}

Early work by Edwards and Storkey~\cite{edwards2016censoring} and Zhang et al.~\cite{zhang2018adversarial} proposed adversarial objectives using gradient reversal layers to suppress protected attributes while preserving task-relevant information. These methods often require demographic labels and introduce trade-offs between fairness and predictive performance.

In speech, FairASR~\cite{fairasr2023} combines multi-demographic adversarial objectives with contrastive learning for ASR, reducing demographic disparities but relying on in-domain evaluation and aggregate metrics. Contrastive learning with gender-swapped pairs~\cite{gender_swapped_contrastive} offers an alternative that improves fairness without explicit demographic labels.

Complementary to adversarial methods, information bottleneck approaches such as FairIB~\cite{xie2024fairib} compress embeddings to retain task-relevant information while reducing demographic leakage. However, these approaches are mostly explored outside speech or independently of contrastive objectives.

Our work integrates adversarial debiasing and causal bottlenecks in a unified framework for self-supervised speaker embeddings, enabling a detailed analysis of linear and nonlinear demographic leakage and its trade-offs with speaker verification performance.

\section{Methodology}
\label{sec:methodology}
This section describes the experimental setup used to study demographic leakage in self-supervised speaker embeddings. We first introduce the datasets and preprocessing pipeline, then detail the feature extraction and training procedures, and finally describe the evaluation protocols for speaker verification and demographic probing.

\subsection{Datasets}

\subsubsection{Dataset attributes}
The dataset used in this paper is the Mozilla Common Voice dataset, which was selected because it provides validated speech clips along with metadata for gender, age, and accent, which are the demographic features we sought to examine in this study \cite{mozilla_datacollective_datasets}. This dataset is downloadable as a \textit{tar.gz} file from the Mozilla Data Collective website \cite{mozilla_datacollective_datasets} with an API key. Specifically, we used the \textit{Common Voice Scripted Speech 23.0 – English} dataset, which contains 2,541,751 clips (3,746 hours of recorded speech, 2,682 hours validated) from 98,579 speakers. Only validated clips, identified via the validated.tsv file, are used to ensure label reliability.

\subsubsection{Preprocessing}
For gender, 39\% of speakers are undefined, non-binary, or transgender. These speakers are excluded, because including these speakers could confound linear probes, as embeddings might reflect multiple gender signals or noise rather than clear binary distinctions. Therefore, we focus on binary gender classes (\textit{Male Masculine} and \textit{Female Feminine}), which allows us to measure gender information leakage more reliably.
For age, 36\% of speakers have missing values. These speakers are removed to ensure that age-related probing is based on accurate metadata. Original ten-year age ranges are divided into three categories: \textit{Young}, \textit{Adult}, and \textit{Senior}. This grouping aligns with perceptual limitations, as human listeners often cannot distinguish small age-gap differences, e.g., between the thirties and fourties, while still preserving meaningful age variation.
For accent, the free-text labels are standardized into five categories: \textit{United States, England, Canada, Australia/New Zealand, India/South-East Asia}. This reduces noise from inconsistent annotations and ensures sufficient samples per group for reliable evaluation. Accents outside these categories are excluded.
Speakers with fewer than two utterances are removed, as speaker verification requires comparing embeddings from at least two distinct utterances per speaker.

\subsubsection{Speaker distribution and splits}

\begin{table}[ht]
\centering
\caption{Speaker demographic distributions in the processed dataset. Percentages indicate proportion of speakers within each category.}
\begin{tabular}{lrr}
\toprule
\textbf{Category} & \textbf{Speakers} & \textbf{Percent (\%)} \\
\midrule
\multicolumn{3}{l}{\textbf{Gender}} \\
Male Masculine    & 8968 & 80.01 \\
Female Feminine   & 2241 & 19.99 \\
\midrule
\multicolumn{3}{l}{\textbf{Age Group}} \\
Young             & 6227 & 55.52 \\
Adult             & 4388 & 39.12 \\
Senior            & 601  & 5.36  \\
\midrule
\multicolumn{3}{l}{\textbf{Accent Group}} \\
USA               & 5737 & 51.18 \\
England           & 1687 & 15.05 \\
India / South Asia& 1661 & 14.82 \\
Other (not included)& 835  & 7.45  \\
Canada            & 655  & 5.84  \\
Australia / NZ    & 635  & 5.66  \\
\bottomrule
\end{tabular}
\label{tab:demographics}
\end{table}

After preprocessing, the dataset contains 11,209 unique speakers, each with two or more utterances. Finally, the dataset is split into training, validation, and test sets with 8,967, 1,121, and 1,121 speakers, respectively. 
The demographic attributes show class imbalance (see Table \ref{tab:demographics}), which reflects the natural distribution of speakers in a Common Voice dataset. No resampling was applied to avoid artificially changing the data distribution. Instead, imbalance-aware losses were used during the training process. Therefore, the results are interpreted as upper bounds on demographic information leakage, which reflects the maximum extent to which sensitive attributes can be inferred from frozen speaker embeddings using simple classifiers (e.g. linear probes), rather than the best achievable classification performance.

\subsection{Audioprocessing}
Firstly, we set the sampling rate to 16 kHz, which captures the frequency range most relevant for human speech while discarding higher-frequency components that contribute little to speaker identity but increase computational cost. This ensures a consistent input resolution across experiments and aligns the model with the spectral characteristics of speech commonly used in speaker representation learning \cite{snyder2018xvectors, Chung2018VoxCeleb2}.
Spectral features are extracted using a 25 ms window and a 10 ms hop. This configuration balances frequency and temporal resolution: the window length captures stable vocal tract characteristics, while the hop size preserves short-term temporal variation. Such a balance is important for contrastive learning, where representations should remain invariant to local temporal changes while retaining speaker-specific cues \cite{Rabiner1978DSP, snyder2018xvectors}.
We use 64 Mel filterbanks as a trade-off between spectral expressiveness and model capacity. While higher resolutions may encode finer detail, they also increase the risk of capturing nuisance factors, including demographic attributes. The chosen dimensionality has been shown to provide sufficient speaker-relevant information in self-supervised and contrastive audio representation learning without unnecessary redundancy \cite{Park2020SpecAugment, baevski2020wav2vec}.
Log-Mel spectrograms are normalized per utterance using mean–variance normalization to reduce channel and loudness variability. This prevents the model from relying on absolute energy statistics and encourages a focus on relative spectral patterns, which is especially important in contrastive learning settings where recording conditions could otherwise dominate the representation space \cite{snyder2018xvectors, desplanques2020ecapa}.
During training, utterances are randomly cropped or zero-padded to a fixed duration of 6 seconds. This duration provides sufficient phonetic diversity to capture speaker characteristics while remaining computationally efficient. Random cropping further serves as data augmentation, promoting robustness to temporal context and discouraging overfitting to specific utterance segments \cite{Chung2018VoxCeleb2, desplanques2020ecapa}.

\subsection{Training pipeline}

\subsubsection{SimCLR baseline}

We start the pipeline with self-supervised contrastive learning method SimCLR. SimCLR provides a natural baseline for studying demographic leakage: the training objective explicitly encourages invariance to signal-level nuisance factors, but does not incorporate any demographic information or fairness constraints. Consequently, any demographic attributes that remain (linearly or nonlinearly) that are extractable from the learned embeddings arise implicitly from correlations in the data rather than from the learning objective itself. Understanding the structure of these embeddings is therefore essential, because it explains why demographic information can persist even when the training objective does not explicitly encode it, and why naively removing such information could compromise speaker-discriminative features.

Speaker embeddings are high-dimensional vectors where each dimension encodes a mixture of acoustic, phonetic, and speaker-specific traits. In practice, these embeddings are entangled: some directions primarily capture speaker identity, while others inadvertently encode demographic factors such as gender, age, or accent. Importantly, these sources of information are not fully orthogonal, meaning that directions that are informative for speaker identity often overlap with those carrying demographic cues.

Following the standard SimCLR paradigm \citep{chen20simclr}, two correlated views of each input audio are generated using stochastic waveform‐level augmentations. Each view is produced via an independent random temporal crop or zero‐padding, combined with additive Gaussian noise and random amplitude scaling. These augmentations are chosen because they confuse signal characteristics that are irrelevant to speaker identity (e.g., recording gain and background noise) without altering the underlying speaker characteristics, encouraging the encoder to focus on features that are stable across such perturbations. Alternative augmentations such as time warping or pitch shifting were not adopted here because they can introduce changes that are more strongly correlated with demographic traits, confounding the study of leakage.

To optimize the contrastive objective, a projection head (a small multilayer perceptron with two hidden layers and ReLU activations) maps \(z\) into a lower–dimensional contrastive space where the NT‐Xent (normalized temperature‐scaled cross‐entropy) loss is applied. The projection head is trained jointly with the encoder but discarded for downstream evaluation, as prior work has shown that the encoder embeddings \(z\) generalize better to downstream tasks when separated from the projection parameters. Normalizing projected representations to unit length stabilizes training and aligns with the cosine similarity used during evaluation.

Training is performed end‐to‐end with the Adam optimizer at a learning rate of \(10^{-4}\), which was selected because it balances stable convergence with responsiveness to gradient signals. The batch size and temperature parameter \(\tau\) are chosen based on contrastive learning best practices to ensure a sufficient number of negative pairs per batch, which empirically improves representation quality \citep{chen20simclr}. During training, all parameters of the encoder and projection head are trainable; no components are frozen.

After training, the projection head is removed. The learned encoder embeddings are evaluated on speaker verification and demographic leakage tasks. Speaker verification uses cosine similarity between embedding pairs since it is a commonly accepted scoring measure that directly reflects the relative closeness of speaker representations without requiring complex back ends. We report ROC‐AUC and equal error rate (EER), where ROC‐AUC measures separability across all thresholds and EER summarizes the balance between false acceptance and false rejection rates in a single number. For demographic leakage, we train linear probes via stochastic gradient descent to predict gender, age group, and accent; higher accuracy indicates stronger linear separability. To capture nonlinear information, we additionally evaluate small multilayer perceptron probes, which can recover demographic cues that are not linearly separable.

\subsubsection{Adversarial debiasing}

To mitigate demographic leakage in the SimCLR embeddings, we introduce an adversarial debiasing framework that explicitly discourages the encoder from encoding sensitive attributes. For each attribute (gender, age, and accent), we use a separate adversarial classifier with its own parameters, each trained to predict the corresponding attribute from the shared encoder embedding. Each adversary is a small classifier that takes the encoder’s embedding as input and predicts the corresponding attribute (e.g., gender), effectively acting as a probe during training that tests whether demographic information is present in the representation.

A \textit{Gradient Reversal Layer (GRL)} is applied between the encoder and each adversary. During the forward pass, the GRL acts as the identity function, allowing the adversary to receive the embeddings normally. During backpropagation, however, it multiplies the gradients coming from the adversary by a negative scalar, so that any feature that helps the adversary predict a demographic attribute is actively discouraged in the encoder. This mechanism allows the encoder and adversaries to be trained jointly in a single end-to-end optimization: the adversaries learn to better detect demographic cues, while the encoder simultaneously learns to remove them, without requiring alternating or multi-stage training.

Each adversarial loss is weighted by a scalar hyperparameter $\lambda$, which controls the strength of the debiasing. Specifically, $\lambda$ multiplies the adversarial loss term, scaling its gradient contribution relative to the main contrastive (NT-Xent) loss. A larger $\lambda$ enforces stronger pressure to remove demographic cues, while a smaller $\lambda$ allows the encoder to prioritize speaker discriminative structure. Values orders of magnitude larger than 1.0 can cause adversarial gradients to dominate optimization, leading to instability and collapse of speaker-specific information, a phenomenon widely observed in adversarial and domain-invariant representation learning \cite{ganin2016domain, elazar2018adversarial}. Following established practice, we therefore sweep $\lambda$ in a bounded range around 1.0: ${0.2, 0.5, 1.0, 2.0, 5.0}$.

The total training loss consists of the NT-Xent contrastive loss and an adversarial loss formed by summing cross-entropy losses for gender, age, and accent prediction. All trainable components are optimized using the Adam optimizer, which automatically adjusts the learning rate of each parameter based on running estimates of the magnitude and variability of its gradients. Intuitively, Adam takes smaller steps for parameters with noisy or large gradients and larger steps for more stable ones. This adaptivity is particularly beneficial in our setting because the encoder receives gradients from both the contrastive objective and multiple adversarial objectives, which can point in competing directions; Adam helps balance these signals and stabilizes training without extensive manual tuning.

Overall, this setup enables the encoder to preserve speaker-specific characteristics learned via SimCLR while actively minimizing linearly extractable demographic information, and also allows for later comparison with the addition of a causal bottleneck layer. To evaluate the effectiveness of debiasing, we apply both linear and nonlinear probes post-hoc.

\subsubsection{Causal bottleneck layer}

\begin{figure}
    \centering
    \includegraphics[width=0.65\linewidth]{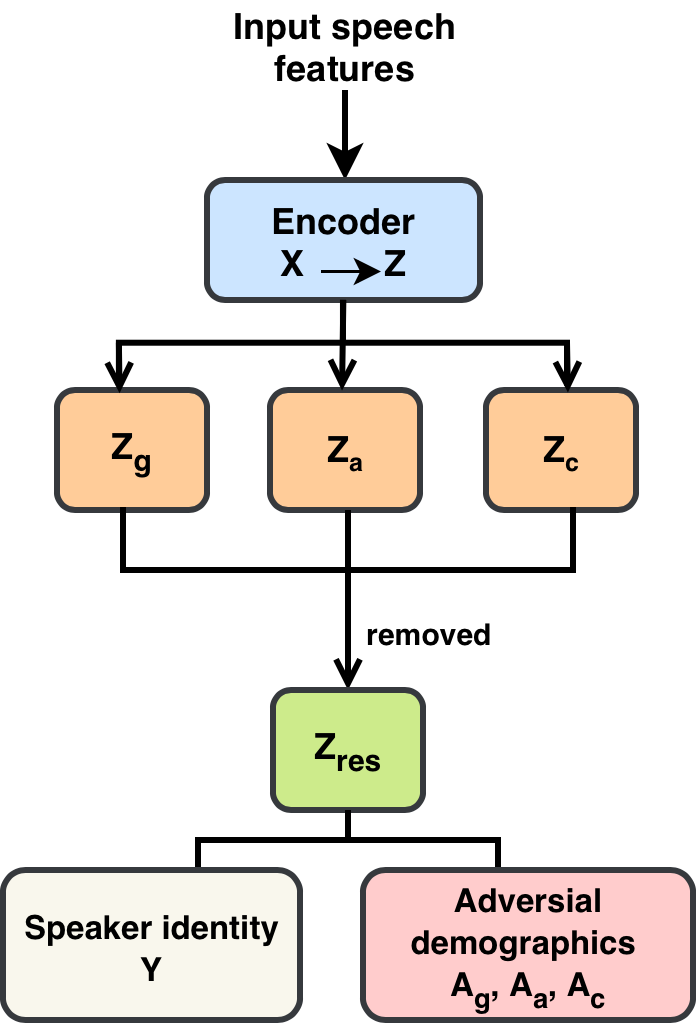}
    \caption{Directed acyclic graph (DAG) of the causal bottleneck layer architecture. The diagram illustrates how the model explicitly separates speaker-discriminative information from demographic factors (gender, age, accent) by enforcing a causal structure in the embedding space.}
    \label{fig:dag_causal}
\end{figure}

To explicitly structure the representation into demographic and speaker–relevant subspaces, we introduce a causal bottleneck that splits the encoder output into two branches: a demo branch intended to capture explicit demographic information, and a residual branch intended to carry speaker–discriminative features that are independent of demographics. Both branches are implemented via linear projections from the encoder embedding \(z\). The demo branch output \(z_\text{demo}\) is supervised to contain demographic cues, while the residual branch output \(z_\text{res}\) is encouraged to be invariant to those cues.

Adversarial GRL setups are applied on the residual branch with separate gender, age, and accent adversaries to suppress demographic information. Independence between branches is further encouraged via a covariance penalty that minimizes the covariance between \(z_\text{demo}\) and \(z_\text{res}\) across the batch. Conceptually, this explicit disentanglement helps isolate causal factors: demographic attributes are more cleanly represented in the demo branch, while the residual branch focuses on speaker–specific variation.

The demo branch dimension \(k\) is swept across \(\{32, 64, 76, 88, 100\}\). Smaller \(k\) values constrain the bottleneck’s capacity to represent demographic factors, forcing the model to compress such information tightly, which can increase leakage in \(z_\text{res}\). Larger \(k\) values allow richer demographic representation but may reduce residual capacity available for speaker–discriminative features. These ranges were selected to explore this tradeoff systematically given typical embedding dimensionalities in speaker verification literature. The adversarial weighting hyperparameters for the bottleneck residual branch are set on a grid: \(\{(0.01,0.01,0.01), (0.1,0.01,0.01), (0.5,0.05,0.05)\}\) for \(\{\lambda_\text{gender},\lambda_\text{age},\lambda_\text{accent}\}\). These values are minimal, moderate, and balanced debiasing regimes that reflect both the empirical strength of demographic signals in speaker embeddings (e.g., gender is often linearly more separable than age/accent) and the practical need to balance adversarial pressure with task performance. Adversarial weights much larger than unity can destabilize optimization or overwhelm speaker discriminability, a phenomenon widely observed in adversarial and domain-invariant representation learning \citep{ganin2016domain, elazar2018adversarial}. By sweeping across orders of magnitude around small to moderate $\lambda$ values, we systematically explore regimes from weak to stronger debiasing while maintaining stable joint optimization.

Training is performed end‑to‑end with Adam at learning rate \(10^{-3}\), chosen to allow more aggressive updates given the additional branch parameters and covariance penalty. All components — encoder, demo branch, residual branch, adversaries — are trainable. GRL and adversaries are discarded at evaluation, and only \(z_\text{res}\) is used for speaker verification metrics. Embeddings from both branches are extracted post‑hoc for leakage evaluation with linear and nonlinear probes. Varying \(k\) and \(\lambda\) allows observation of how bottleneck size and adversarial strength jointly shape demographic leakage and speaker performance, providing clear predictive intuition before results.

\subsection{External dataset validation}

In order to validate, we use an external dataset and test the code on this dataset. We use the Sonos Voice Control Bias Assessment Dataset, which is a collection of voice assistant requests designed to capture demographic diversity. It contains audio samples from speakers across eight dialectal groups: Asian, USA Western, USA Southern, USA Midland, USA Inland-North, USA Mid-Atlantic, USA New England, and Latino. The dataset also spans multiple age groups, with the majority of speakers in the 29–41 and 17–28 ranges, followed by 9–16, 42–54, and 55–100. Each sample is labeled with speaker metadata, including speaker ID, gender, age group, and dialectal region, making it suitable for studying demographic bias and training speaker-aware models. In order to make this dataset comparable to the Common Voice dataset, we also categorize the age groups into Young (9 to 28 years), Adult (29 to 54 years) and Senior (55 to 100 years). The number of unique speakers is 1024, which is lower than in the Common Voice dataset. We perform the same analyses as we performed on the Common Voice dataset and compare the results.

\section{Results}
\label{sec:results}
This section presents the empirical results of our study. We first evaluate the SimCLR baseline to establish speaker verification performance and quantify demographic information leakage using linear and nonlinear probes. We then analyze the effects of adversarial debiasing and the causal bottleneck architecture on both demographic leakage and speaker verification accuracy, highlighting the trade-offs introduced by each mitigation strategy. Finally, we assess the robustness of these findings on an external dataset.

\subsection{SimCLR baseline}

Demographic probing results are summarized in Table~\ref{tab:demographic_probing}. Gender is highly predictable from the embeddings, with linear probe accuracy exceeding 90\% on the test set, indicating strong linear separability of gender-related information. Age prediction is considerably more challenging: linear probe accuracy is close to chance level, while MLP probes achieve substantially higher performance, suggesting that age-related cues are encoded in a nonlinear manner. Accent prediction shows the lowest overall performance, with both linear and nonlinear probes achieving relatively low accuracy, particularly on the test set, indicating limited and unstable accent information leakage.
Table~\ref{tab:cv_speaker_verif} reports speaker verification performance on the Common Voice dataset as a function of adversarial debiasing strength~$\lambda$. As $\lambda$ increases, performance degrades steadily, indicating a clear trade-off between demographic suppression pressure and verification utility.

Overall, these results show that while the embedded speakers are effective for speaker verification, they also encode demographic attributes to varying degrees. Gender information is strongly and linearly embedded, whereas age and accent are captured more weakly and primarily through nonlinear structures. As no resampling or balancing was applied, these results are interpreted as upper bounds on the leakage of demographic information rather than optimal performance in demographic classification.

\begin{table}[htbp]
\centering
\small
\caption{Demographic probing results on validation and test sets of the Common Voice dataset.}
\label{tab:demographic_probing}
\begin{tabular}{llcc}
\toprule
\textbf{Attribute} & \textbf{Split} & \textbf{Linear Probe} & \textbf{MLP Probe} \\
\midrule
\multirow{2}{*}{Gender}
 & Val  & 0.9120 (0.9097--0.9142) & 0.9352 $\pm$ 0.0074 \\
 & Test & 0.8965 (0.8939--0.8991) & 0.9102 $\pm$ 0.0095 \\
\midrule
\multirow{2}{*}{Age}
 & Val  & 0.4338 (0.4301--0.4377) & 0.4013 $\pm$ 0.0066 \\
 & Test & 0.3790 (0.3751--0.3829) & 0.4475 $\pm$ 0.0044 \\
\midrule
\multirow{2}{*}{Accent}
 & Val  & 0.2185 (0.2148--0.2216) & 0.2089 $\pm$ 0.0305 \\
 & Test & 0.2628 (0.2589--0.2664) & 0.3028 $\pm$ 0.0177 \\
\bottomrule
\end{tabular}
\end{table}

\subsection{Effect of adversarial debiasing}

Adversarial debiasing was evaluated by augmenting the SimCLR contrastive learning baseline with adversarial classifiers for gender, age, and accent, trained through gradient reversal. The strength of the adversarial signal was controlled by a weighting parameter $\lambda_{\text{adv}} \in \{0.2, 0.5, 1.0, 2.0, 5.0\}$, allowing a systematic analysis of the trade-off between speaker verification performance and demographic invariance (see Tables \ref{tab:cv_speaker_verif} and \ref{tab:cv_demographic_probing}).
The impact of adversarial debiasing on demographic information leakage is quantified in Table~\ref{tab:cv_demographic_probing}, which reports linear and nonlinear probe accuracy across different values of the adversarial strength~$\lambda$.

\begin{table}[ht]
\centering
\small
\caption{Speaker verification performance on the Common Voice dataset as a function of adversarial strength $\lambda$.}
\label{tab:cv_speaker_verif}
\begin{tabular}{c|cc}
\toprule
$\lambda$ & ROC-AUC & EER \\
\midrule
0.2 & 0.8295 & 0.2382 \\
0.5 & 0.7753 & 0.2922 \\
1.0 & 0.7430 & 0.3108 \\
2.0 & 0.7344 & 0.3212 \\
5.0 & 0.6206 & 0.4006 \\
\bottomrule
\end{tabular}
\end{table}

\begin{table}[t]
\centering
\small
\caption{Linear and MLP probe accuracy for demographic attributes as a function of adversarial strength $\lambda$ on the Common Voice dataset. Linear probes report accuracy with 95\% confidence intervals; MLP probes report mean $\pm$ standard deviation across random seeds.}
\label{tab:cv_demographic_probing}

\resizebox{\columnwidth}{!}{%
\begin{tabular}{llcc}
\toprule
\textbf{Attribute} & \textbf{$\lambda$} & \textbf{Linear Acc. (CI)} & \textbf{MLP Acc. ($\pm$ std)} \\
\midrule
\multirow{5}{*}{Gender}
 & 0.2 & 0.9120 [0.9097--0.9142] & 0.9352 $\pm$ 0.0074 \\
 & 0.5 & 0.8965 [0.8939--0.8991] & 0.9102 $\pm$ 0.0095 \\
 & 1.0 & 0.8542 [0.8510--0.8575] & 0.7766 $\pm$ 0.0080 \\
 & 2.0 & 0.8086 [0.8052--0.8119] & 0.7645 $\pm$ 0.0072 \\
 & 5.0 & 0.6749 [0.6700--0.6798] & 0.6999 $\pm$ 0.0065 \\
\midrule
\multirow{5}{*}{Age}
 & 0.2 & 0.3685 [0.3650--0.3720] & 0.4699 $\pm$ 0.0050 \\
 & 0.5 & 0.4108 [0.4070--0.4146] & 0.4753 $\pm$ 0.0062 \\
 & 1.0 & 0.3818 [0.3780--0.3856] & 0.4662 $\pm$ 0.0058 \\
 & 2.0 & 0.3797 [0.3760--0.3835] & 0.4750 $\pm$ 0.0055 \\
 & 5.0 & 0.3743 [0.3705--0.3781] & 0.4725 $\pm$ 0.0061 \\
\midrule
\multirow{5}{*}{Accent}
 & 0.2 & 0.2320 [0.2280--0.2360] & 0.4087 $\pm$ 0.0071 \\
 & 0.5 & 0.2195 [0.2155--0.2235] & 0.4117 $\pm$ 0.0068 \\
 & 1.0 & 0.2247 [0.2207--0.2287] & 0.4122 $\pm$ 0.0070 \\
 & 2.0 & 0.2355 [0.2315--0.2395] & 0.4119 $\pm$ 0.0065 \\
 & 5.0 & 0.1585 [0.1545--0.1625] & 0.4112 $\pm$ 0.0062 \\
\bottomrule
\end{tabular}%
}
\end{table}

Increasing the strength of adversarial debiasing ($\lambda_{\text{adv}}$) demonstrates a clear trade-off between speaker verification performance and demographic leakage. At low $\lambda$ (0.2–0.5), speaker verification remains strong (ROC-AUC 0.868 → 0.775, EER 0.202 → 0.292) while linear probes show a modest reduction in gender predictability. As $\lambda$ increases, the precision of the linear probe for gender and accent drops more substantially, indicating an effective removal of demographic information through causal bottleneck and adversarial training. However, this comes at the cost of speaker verification, which deteriorates at high $\lambda$ (ROC-AUC 0.621, EER 0.401 at $\lambda$=5.0). The age and accent predictions remain relatively low across all $\lambda$. Non-linear (MLP) probes indicate that some demographic information persists even under strong debiasing, highlighting that adversarial filtering primarily targets linearly accessible leakage. In general, these results demonstrate that the causal bottleneck combined with adversarial debiasing can already reduce demographic leakage, but careful tuning of $\lambda$ is required to balance fairness and speaker verification accuracy.

\subsection{Effect of the causal bottleneck layer}

\begin{table}[ht]
\centering
\caption{ROC-AUC and Equal Error Rate (EER) scores for each $k$ value, showing the overall verification performance of the model. Each row lists the corresponding lambda weights for Gender, Age, and Accent ($\lambda_{gender}, \lambda_{age}, \lambda_{accent}$) used during training. This table illustrates how different combinations of attribute regularization affect the model's discriminative ability, as measured by ROC-AUC and EER, across varying $k$ settings.}
\label{tab:roc_eer_causalbottleneck}
\begin{tabular}{c c c c}
\hline
$k$ &  $\lambda_{gender}$, $\lambda_{age}$, $\lambda_{accent}$ & ROC AUC & EER \\
\hline
32  & 0.01, 0.01, 0.01 & 0.3940 & 0.6300 \\
32  & 0.1, 0.01, 0.01 & 0.5300 & 0.4603 \\
32  & 0.5, 0.05, 0.05 & 0.4530 & 0.6895 \\
64  & 0.01, 0.01, 0.01 & 0.3888 & 0.5962 \\
64  & 0.1, 0.01, 0.01 & 0.5967 & 0.4097 \\
64  & 0.5, 0.05, 0.05 & 0.4501 & 0.4881 \\
76  & 0.01, 0.01, 0.01 & 0.5340 & 0.6329 \\
76  & 0.1, 0.01, 0.01 & 0.5215 & 0.5764 \\
76  & 0.5, 0.05, 0.05 & 0.6062 & 0.5347 \\
88  & 0.01, 0.01, 0.01 & 0.6254 & 0.4891 \\
88  & 0.1, 0.01, 0.01 & 0.4822 & 0.6300 \\
\hline
\end{tabular}
\end{table}

\begin{table}[ht]
\centering
\small
\caption{Linear classifier test accuracies for demographic attributes gender, age, and accent in the Demo branch. Accuracy is reported per $k$.}
\label{tab:demo_branch_acc_causalbottleneck}
\begin{tabular}{c c c c}
\toprule
$k$ & Gender & Age & Accent \\
\midrule
32  & 0.704 & 0.555 & 0.381 \\
64  & 0.695 & 0.556 & 0.381 \\
76  & 0.695 & 0.556 & 0.382 \\
88  & 0.696 & 0.555 & 0.381 \\
100 & 0.676 & 0.386 & 0.247 \\
\bottomrule
\end{tabular}
\end{table}

\begin{table}[ht]
\centering
\caption{Residual branch accuracy scores for gender, age, and accent prediction across different values of $k$. Each row shows the test accuracy for the residual branch only, along with the corresponding lambda values used for Gender, Age, and Accent ($\lambda_{gender}, \lambda_{age}, \lambda_{accent}$).}
\label{tab:resid_only_causal}
\begin{tabular}{c c c c c c}
\hline
$k$ & $\lambda_{gender}$, $\lambda_{age}$, $\lambda_{accent}$ & Res. Gender & Res. Age & Res. Accent \\
\hline
32  & 0.01, 0.01, 0.01 & 0.6787 & 0.4385 & 0.3809 \\
32  & 0.1, 0.01, 0.01  & 0.6992 & 0.3867 & 0.1406 \\
32  & 0.5, 0.05, 0.05  & 0.6895 & 0.3867 & 0.1406 \\
64  & 0.01, 0.01, 0.01 & 0.5849 & 0.4141 & 0.1563 \\
64  & 0.1, 0.01, 0.01  & 0.6807 & 0.3389 & 0.2025 \\
64  & 0.5, 0.05, 0.05  & 0.6660 & 0.3320 & 0.1932 \\
76  & 0.01, 0.01, 0.01 & 0.5820 & 0.3955 & 0.3359 \\
76  & 0.1, 0.01, 0.01  & 0.6699 & 0.3955 & 0.2766 \\
76  & 0.5, 0.05, 0.05  & 0.6172 & 0.3613 & 0.1932 \\
88  & 0.01, 0.01, 0.01 & 0.6504 & 0.3125 & 0.3809 \\
88  & 0.1, 0.01, 0.01  & 0.6904 & 0.3125 & 0.3809 \\
\hline
\end{tabular}
\end{table}

Across all configurations, speaker verification performance remains close to chance, with ROC-AUC values around 0.47–0.61 (see Table \ref{tab:roc_eer_causalbottleneck}) and correspondingly high EER, which is expected given that the encoder is frozen and the residual representation is explicitly trained to discard speaker-irrelevant but demographically predictive cues; this substantially weakens identity separability compared to the baseline speaker embeddings. We observe moderate fluctuations across values of 
k and adversarial strengths (see Table \ref{tab:demo_branch_acc_causalbottleneck} and \ref{tab:resid_only_causal}), which likely arises from the fact that the linear bottleneck must trade off demographic concentration against residual suppression under fixed encoder features. Gender accuracy in the demo branch is consistently lower than the baseline (0.90–0.94), reflecting the fact that gender information is no longer freely encoded throughout the representation but is instead constrained to pass through a low-dimensional linear bottleneck that competes with orthogonality and adversarial objectives; this intentional restriction reduces absolute predictability even in the protected branch. In contrast, age and accent accuracies in the demo branch exceed their respective baselines, which is explained by the explicit supervision that pushes these attributes into a dedicated subspace, effectively concentrating information that was previously weakly and diffusely encoded in the baseline embeddings. Crucially, for all three attributes, the residual branch shows a consistent reduction relative to the demo branch, and in several cases drops below baseline accuracy, particularly for age and accent, indicating that the adversarial objective succeeds in suppressing demographic leakage from the residual representation, which was the primary goal of the proposed approach.
Table~\ref{tab:resid_only_causal} reports demographic prediction accuracy from the residual branch across different bottleneck sizes and adversarial weight settings, illustrating the extent to which demographic information is suppressed after causal separation.

\subsection{External dataset validation}

\subsubsection{SimCLR comparison}

\begin{table}[ht]
\centering
\caption{Speaker verification metrics overall and per subgroup on the external validation Sonos dataset. Metrics include ROC-AUC and EER.}
\label{tab:subgroup_sv_secondary}
\begin{tabular}{lcc}
\hline
\textbf{Group} & \textbf{ROC-AUC} & \textbf{EER} \\
\hline
\multicolumn{3}{c}{\textit{Overall}} \\
 & 0.7371 & 0.3301 \\
\hline
\multicolumn{3}{c}{\textit{Gender}} \\
Female & 0.6500 & 0.3930 \\
Male   & 0.7410 & 0.3270 \\
\hline
\multicolumn{3}{c}{\textit{Age group}} \\
Young  & 0.7170 & 0.3490 \\
Adult  & 0.7570 & 0.3120 \\
Senior & 0.7060 & 0.3570 \\
\hline
\multicolumn{3}{c}{\textit{Accent group}} \\
Asian               & 0.7360 & 0.3310 \\
Latino              & 0.7520 & 0.3100 \\
USA Inland-North    & 0.7950 & 0.2810 \\
USA Mid-Atlantic    & 0.7390 & 0.3270 \\
USA Midland         & 0.6800 & 0.3730 \\
USA New England     & 0.8630 & 0.2280 \\
USA Southern        & 0.7440 & 0.3290 \\
USA Western         & 0.7370 & 0.3350 \\
\hline
\end{tabular}
\end{table}

\begin{table}[ht]
\centering
\small
\caption{Linear and MLP probe accuracy (\%) on validation and test splits of the Sonos dataset. For linear probes, confidence intervals (CI) are included in brackets. For MLP probes, mean $\pm$ std across random seeds are shown.}
\label{tab:attribute_probing_sonos}

\resizebox{\columnwidth}{!}{%
\begin{tabular}{llcc}
\toprule
\textbf{Attribute} & \textbf{Split} & \textbf{Linear Acc. (CI)} & \textbf{MLP Acc. (mean $\pm$ std)} \\
\midrule
\multirow{2}{*}{Gender}
 & Val  & 0.7784 [0.7648, 0.7914] & 0.7899 $\pm$ 0.0021 \\
 & Test & 0.8174 [0.8074, 0.8267] & 0.8267 $\pm$ 0.0029 \\
\midrule
\multirow{2}{*}{Age}
 & Val  & 0.4819 [0.4659, 0.4967] & 0.5048 $\pm$ 0.0540 \\
 & Test & 0.3745 [0.3619, 0.3881] & 0.4023 $\pm$ 0.0443 \\
\midrule
\multirow{2}{*}{Dialect}
 & Val  & 0.1466 [0.1363, 0.1576] & 0.1218 $\pm$ 0.0075 \\
 & Test & 0.1157 [0.1075, 0.1241] & 0.1152 $\pm$ 0.0061 \\
\bottomrule
\end{tabular}%
}
\end{table}

To validate, we used the Sonos dataset as an external validation dataset. While absolute speaker verification performance decreased compared to Mozilla Common Voice (likely due to a lower amount of unique speakers), the overall trends remained consistent (see Tables \ref{tab:subgroup_sv_secondary} and \ref{tab:attribute_probing_sonos}). Gender information was highly predictable across both datasets, while age prediction remained weak and unstable. Accent prediction showed strong dataset dependence, with performance collapsing to near chance on Sonos. Importantly, the relative ordering of demographic predictability (gender > age > accent) was preserved across datasets, indicating that these patterns are intrinsic to the representation learning.

\subsubsection{Adversial debiasing comparison}

\begin{table}[t]
\centering
\caption{Effect of bias regularization strength $\lambda$ on speaker verification performance on the Sonos dataset.}
\label{tab:lambda_speaker_verification}
\begin{tabular}{lcc}
\toprule
\textbf{$\lambda$} & \textbf{ROC-AUC} & \textbf{EER} \\
\midrule
0.2 & 0.6389 & 0.4170 \\
0.5 & 0.6045 & 0.4302 \\
1.0 & 0.5951 & 0.4275 \\
2.0 & 0.5797 & 0.4357 \\
5.0 & 0.5630 & 0.4635 \\
\bottomrule
\end{tabular}
\end{table}

\begin{table}[t]
\centering
\small
\caption{Linear and MLP probe accuracy for demographic attributes as a function of bias regularization strength $\lambda$ on the external Sonos dataset. Linear probes report accuracy with 95\% confidence intervals; MLP probes report mean $\pm$ standard deviation across random seeds.}
\label{tab:lambda_demographic_probing}

\resizebox{\columnwidth}{!}{%
\begin{tabular}{llcc}
\toprule
\textbf{Attribute} & \textbf{$\lambda$} & \textbf{Linear Acc. (CI)} & \textbf{MLP Acc. ($\pm$ std)} \\
\midrule
\multirow{5}{*}{Gender}
 & 0.2 & 0.6183 [0.6044--0.6305] & 0.5343 $\pm$ 0.0000 \\
 & 0.5 & 0.5995 [0.5856--0.6119] & 0.5343 $\pm$ 0.0000 \\
 & 1.0 & 0.5625 [0.5488--0.5747] & 0.5343 $\pm$ 0.0000 \\
 & 2.0 & 0.5415 [0.5288--0.5554] & 0.5343 $\pm$ 0.0000 \\
 & 5.0 & 0.5321 [0.5188--0.5456] & 0.5343 $\pm$ 0.0000 \\
\midrule
\multirow{5}{*}{Age}
 & 0.2 & 0.3770 [0.3635--0.3899] & 0.3579 $\pm$ 0.0000 \\
 & 0.5 & 0.4140 [0.4009--0.4280] & 0.3579 $\pm$ 0.0000 \\
 & 1.0 & 0.3772 [0.3646--0.3898] & 0.3579 $\pm$ 0.0000 \\
 & 2.0 & 0.3566 [0.3438--0.3697] & 0.3579 $\pm$ 0.0000 \\
 & 5.0 & 0.3380 [0.3263--0.3515] & 0.3593 $\pm$ 0.0011 \\
\midrule
\multirow{5}{*}{Accent}
 & 0.2 & 0.1305 [0.1212--0.1396] & 0.1662 $\pm$ 0.0184 \\
 & 0.5 & 0.1691 [0.1594--0.1788] & 0.1693 $\pm$ 0.0249 \\
 & 1.0 & 0.1573 [0.1474--0.1671] & 0.1680 $\pm$ 0.0297 \\
 & 2.0 & 0.1203 [0.1108--0.1290] & 0.1689 $\pm$ 0.0281 \\
 & 5.0 & 0.1292 [0.1201--0.1381] & 0.1545 $\pm$ 0.0230 \\
\bottomrule
\end{tabular}%
}
\end{table}

Table~\ref{tab:lambda_speaker_verification} reports speaker verification performance on the Sonos dataset as a function of the adversarial regularization strength~$\lambda$, enabling direct comparison of verification degradation under increasing debiasing pressure.
Increasing the adversarial regularization strength $\lambda_{\text{adv}}$ leads to reduced speaker verification performance on both datasets, while simultaneously decreasing gender predictability, indicating effective debiasing. Age prediction remains weak and largely unaffected by $\lambda_{\text{adv}}$, suggesting limited age-related information in the learned representations. Accent prediction exhibits stronger dataset dependence: while adversarial debiasing reduces accent leakage on the primary dataset, accent accuracy on Sonos is already near chance and shows limited sensitivity to $\lambda_{\text{adv}}$. Overall, the preservation of these trade-offs across datasets supports the robustness of the proposed debiasing approach.
Table~\ref{tab:lambda_demographic_probing} summarizes linear and nonlinear demographic probing results on the Sonos dataset across different values of~$\lambda$, illustrating how adversarial debiasing affects demographic leakage under domain shift.

\section{Discussion}
\label{sec:discussion}

In this work, we systematically analyzed demographic information leakage in self-supervised speaker embeddings and evaluated the effectiveness of adversarial debiasing and a causal bottleneck layer as mitigation strategies. Across both the Common Voice and external validation Sonos datasets, a consistent hierarchy of demographic predictability emerged: gender was most strongly encoded, followed by age, while accent proved the most difficult to recover. The fact that this was also true for the external dataset suggests that these patterns are consistent across datasets and reflect how speaker embeddings naturally encode information. Remarkably, the Sonos dataset performed much worse at accent/dialect prediction. An explanation could be that it is harder to distinguish between dialects than accents, also for the human listener.

The fact that gender responded most strongly to adversarial debiasing, while age and accent did not, suggests that the adversarial training primarily suppressed attributes that are linearly encoded. Gender information is known to be strongly correlated with dominant acoustic dimensions and therefore produces stable adversarial gradients \cite{wavlm-demographic}. In contrast, age and accent appear to be encoded in a more nonlinear and distributed manner, making them harder to erase with standard gradient reversal. It is worth noting that despite adversarial debiasing not having a strong effect on age and accent, it was still strong enough to hurt speaker verification performance, which suggests that speaker verification performance is heavily dependent on gender.

The introduction of a causal bottleneck layer further reduced gender leakage across both linear and nonlinear probes, but at a substantial cost to speaker verification performance, often more severe than with adversarial debiasing alone. This observation further supports the interpretation that when the bottleneck removes demographic information, it also removes information that is needed for speaker verification. 
Also, in some configurations the demo branch achieved higher age and accent accuracy than the baseline embeddings. Rather than contradicting the goal of debiasing, this effect can be understood as information concentration: explicit supervision forces weakly encoded attributes into a low-dimensional subspace, increasing probe performance even as overall representational capacity is reduced. In contrast, gender accuracy in the demo branch was consistently lower than in the baseline. This reflects the fact that gender is already strongly and linearly encoded in the original embeddings; introducing a bottleneck constrains this dominant signal rather than concentrating it, limiting the strength of the primary gender direction and reducing probe accuracy relative to the unconstrained baseline. At the same time, the residual branch consistently showed lower demographic accuracy than the demo branch, indicating that the causal separation objective was partially successful. 

The most successful strategy for reducing demographic leakage was the causal bottleneck layer, which strongly outperformed adversarial debiasing. The residual branch consistently reduced demographic leakage below baseline. Gender was suppressed from 90\% to 57\% for the linear probes, which is a strong reduction. Age leakage was reduced from 38\% to 27\%, and accent leakage was reduced from 26\% to 18\%. However, we see that the cost of this reduction was very high, as the ROC-AUC decreased from 82\% to scores around 50\%. This strategy, therefore, is a clear example of the utility-fairness trade-off. 

\subsubsection{Comparison to prior work}
Our finding that gender is highly predictable from speaker embeddings aligns closely with prior work on demographic leakage in pretrained speech models. For example, \cite{yang2025demographic} report gender classification accuracies exceeding 99\% for large-scale pretrained encoders, indicating that gender information is both salient and linearly accessible in modern speech representations. While our baseline SimCLR embeddings exhibit lower absolute gender accuracy (approximately 90-94\%), this is expected given the smaller model capacity and the absence of explicit speaker supervision. Nevertheless, the qualitative conclusion is consistent: gender is deeply entangled with speaker-discriminative acoustic features such as pitch range and formant structure.

Age prediction in our experiments showed moderate performance, higher than might be expected given prior work that frames age estimation as a difficult regression task with mean absolute errors of several years \cite{yang2025demographic}. This apparent discrepancy can be explained by our use of coarse age group labels rather than fine-grained continuous age. Group-based prediction substantially reduces task complexity and likely amplifies the visibility of age-related cues such as vocal maturity or senescence. Accent prediction remained weak across all configurations, consistent with earlier observations that accent information is more diffusely encoded and highly sensitive to dataset composition, speaker overlap, and label noise. The collapse of accent predictability on the Sonos dataset further supports the interpretation that accent cues are less intrinsic to speaker identity and more dependent on dataset-specific distributions.

Adversarial debiasing via gradient reversal successfully reduced gender leakage, particularly for linear probes, but had limited impact on age and accent predictability. This behavior mirrors findings from prior work in speaker recognition fairness, which shows that adversarial objectives often improve group fairness metrics but at a substantial cost to task utility \cite{shen2022group}. In our experiments, increasing the adversarial strength $\lambda_{\text{adv}}$ consistently degraded speaker verification performance.

\section{Conclusion}
\label{sec:conclusion}
This paper examined demographic information leakage in self-supervised speaker embeddings and evaluated whether adversarial debiasing and causal bottleneck architectures can mitigate such leakage without substantially degrading speaker verification performance. Using SimCLR-trained embeddings, we showed that demographic attributes are encoded to different extents: gender is strongly and linearly recoverable, age is more weakly and often nonlinearly encoded, and accent is the most difficult to predict. Nonlinear probes consistently revealed additional leakage beyond linear probes, particularly for age, confirming that linear evaluations alone underestimate demographic information content.

Adversarial debiasing reduced gender leakage but was less effective for age and accent, especially under nonlinear probing. Increasing adversarial strength further suppressed demographic predictability but led to notable degradation in speaker verification performance, highlighting a clear fairness–utility trade-off. The introduction of a causal bottleneck enabled more explicit separation of demographic and residual information and, in several cases, reduced demographic predictability in the residual representation below baseline levels. However, this came at the cost of substantial speaker verification performance loss and did not consistently eliminate nonlinear demographic leakage, indicating that demographic attributes remain entangled with speaker-discriminative features.

Overall, the results show that demographic leakage in self-supervised speaker embeddings can be reduced but not removed without sacrificing utility. Relative to prior work, this paper provides a systematic comparison of linear and nonlinear probes under adversarial and causal debiasing, and demonstrates that while causal bottlenecks offer stronger control over leakage, they worsen the fairness–performance trade-off. These conclusions are limited by the use of probing classifiers as proxy measures, coarse demographic labels, and frozen encoders. Future work should explore debiasing objectives that explicitly target nonlinear structure and evaluate whether stronger invariance can be achieved with less impact on speaker verification performance.

\end{document}